\documentclass[journal]{IEEEtran}
\ifCLASSINFOpdf
  \usepackage[pdftex]{graphicx}
  
\else
\fi
\usepackage{amsmath}
\usepackage{amssymb}
\usepackage{xcolor}
\usepackage{booktabs}
\usepackage{multirow}
\usepackage{graphicx}
\usepackage{pifont}
\usepackage{multirow}
\usepackage{color}
\begin{document}
%

\def\baseline{SRCMSA}
\def\etal{et al.}
\def\ourtaskfull{Video Question-Answer Generation}
\def\ourtask{VQAG}
\def\ourgenfull{Joint Question-Answer Generator}
\def\ourgen{JQAG}
\newcommand{\revision}[1]{#1}

\makeatletter

\def\zapcolorreset{\let\reset@color\relax\ignorespaces}
\def\colorrows#1{\noalign{\aftergroup\zapcolorreset#1}\ignorespaces}

\makeatother
\newcommand{\revisionrow}[1]{\colorrows{\color{black}}}

\def\oursfull{Generator-Pretester Network}
%
\def\ours{GPN}
\def\ourlossfull{Target Consistency Loss}
\def\ourloss{TCL}
\def\ourmodule{PT}
\def\ourmodulefull{Pretester}
\def\ouragfull{Answer Proposal Generator}

\title{End-to-End Video Question-Answer Generation with Generator-Pretester Network}
%
%
%
\author{Hung-Ting Su$\dag$, Chen-Hsi Chang$\dag$, Po-Wei Shen$\dag$, Yu-Siang Wang$\ddagger$, \\Ya-Liang Chang$\dag$, Yu-Cheng Chang$\dag$, Pu-Jen Cheng$\dag$ and~Winston H. Hsu$\dag$
\\
$\dag$National Taiwan University, $\ddagger$University of Toronto\\
}
\maketitle

\begin{abstract}
We study a novel task, \ourtaskfull{} (\ourtask{}), for challenging Video Question Answering (Video QA) task in multimedia. Due to expensive data annotation costs, many widely used, large-scale Video QA datasets such as Video-QA, MSVD-QA and MSRVTT-QA are automatically annotated using Caption Question Generation (CapQG) which inputs captions instead of the video itself. 
As captions neither fully represent a video, nor are they always practically available, it is crucial to generate question-answer pairs based on a video via Video Question-Answer Generation (VQAG). Existing video-to-text (V2T) approaches, despite taking a video as the input, only generate a question alone. 
In this work, we propose a novel model \oursfull{} that focuses on two components: (1) The \ourgenfull{} (\ourgen{}) which generates a question with its corresponding answer to allow Video Question ``Answering” training. (2) The \ourmodulefull{} (\ourmodule{}) verifies a generated question by trying to answer it and checks the pretested answer with both the model's proposed answer and the ground truth answer. 
We evaluate our system with the only two available large-scale human-annotated Video QA datasets and achieves state-of-the-art question generation performances. Furthermore, using our generated QA pairs only on the Video QA task, we can surpass some supervised baselines.
As a pre-training strategy, we outperform both CapQG and transfer learning approaches when employing semi-supervised (20\%) or fully supervised learning with annotated data.
These experimental results suggest the novel perspectives for Video QA training.
\end{abstract}

\begin{IEEEkeywords}
video question answering, video question generation, pretester network.
\end{IEEEkeywords}

%
\IEEEpeerreviewmaketitle

\begin{figure*}
  \includegraphics[width=\textwidth]{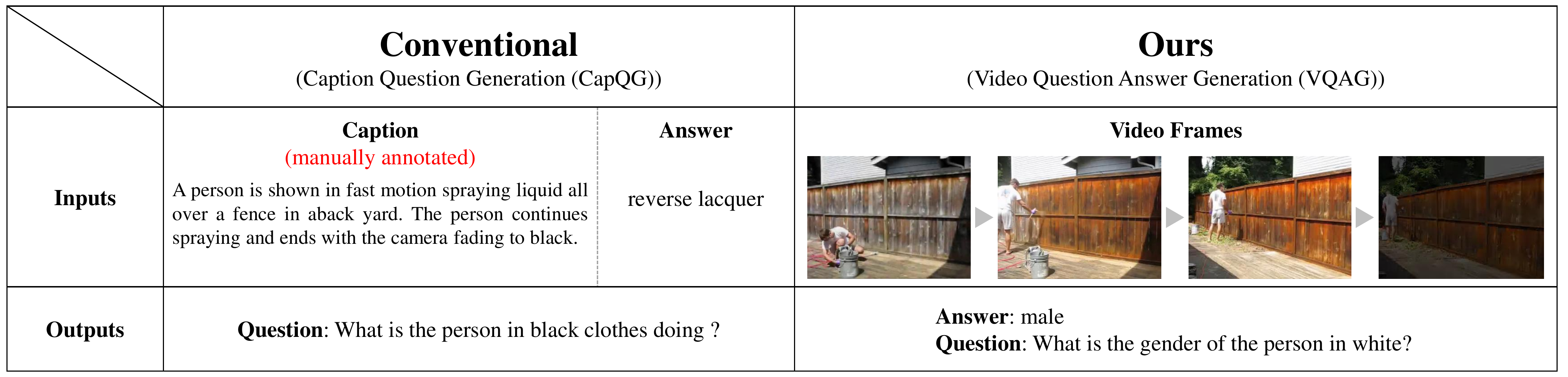}
  \caption{\ourtaskfull{} (\ourtask{}).
  Conventional Caption Question Generation (CapQG, Left column) \cite{vdqa0,vdqa1} which outputs a question with \textit{captions} and an explicitly generated answer. CapQG uses captions which neither fully represent the video, nor are they always practically available. We propose a novel \ourtask{} (Right Column) that automatically generates \textit{question-answer pairs} according to a \textit{video clip} for Video Question Answering (Video QA) training in an end-to-end manner. [Best viewed in color.]
  }
  \label{fig:teaser}
\end{figure*}

\maketitle

\section{Introduction}

Video Question Answering (Video QA; \revision{Table \ref{tab:abbe} lists the abbreviations used in this paper.}), which aims to answer a natural language question according to a video clip, is an important task in multimedia understanding.
Modern Video QA systems \cite{vdqa0,vdqa1,vdqa2,vdqa3,vdqa7,vdqa5,vdqa6,vdqa8,vdqa9,vdqa10,vdqgtmm1,vdqatcvst1}, \revision{\cite{rvdqa2}} require sufficient video-question-answer triples to train. As shown in Figure \ref{fig:teaser}, many Video QA datasets are labeled with \textit{caption question generation (CapQG)}, where question-answer pairs are created using a text question generation system according to captions of a video. 
Three of the most widely used benchmarks, Video-QA \cite{vdqa1}, MSVD-QA, and MSRVTT-QA \cite{vdqa0} are massively labeled in this way and provide more than 50,000 question-answer pairs. 

\begin{figure}
  \includegraphics[width=\columnwidth]{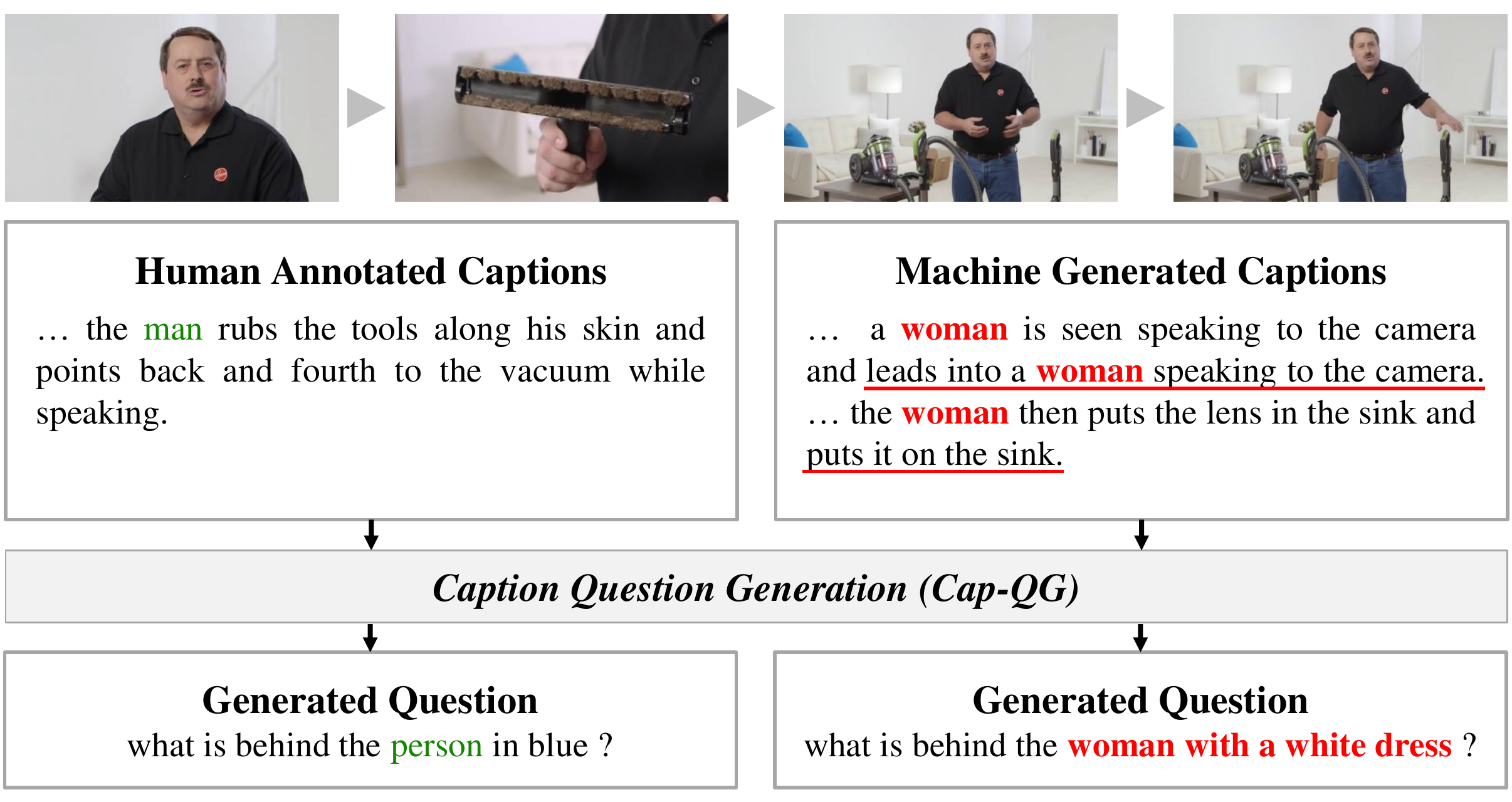}
  \caption{When human annotated captions (Left) are unavailable, applying machine generated captions (Right) to question generation systems incurs \textit{captioning error} and results in \textit{unanswerable questions} generated. 
  Our proposed \ourtask{} directly generates questions from a video and tries to reduce above information loss.
  [Best viewed in color.]}
  \label{fig:dcap}
\end{figure}

Two problems remain unsolved for applying CapQG to real-world applications. First, captions are  
difficult to obtain in practice. Second, these datasets assume captions could represent a video clip, but as the old saying goes, \textit{A picture is worth a thousand words}. Even a single frame has abundant visual information about various objects and interactions between them, and this information increases with the video length. 
An intuitive way to approach videos without captions is to apply video captioning systems. However, video captioning is still an unsolved task that requires deep and fine-grained understanding of a video. 
Although modern systems can generate ``correct” captions in terms of some automatic evaluation metrics such as BLEU, 
qualities of machine generated captions are still not applicable for CapQG due to \textbf{factual errors} where a few erroneous words could lead the captions to have totally different meaning, as shown in Figure \ref{fig:dcap}.
Moreover, correct captions can be inferior because of a general problem such as repetition or redundancy \cite{videocap1}. 
Existing Video QA datasets generated with human-labeled descriptions still suffer from information loss during captioning. Although human annotators can write high-quality captions to a video clip, multimedia contents are too abundant to describe with sentences. For example, Figure \ref{fig:teaser} shows a video and human written captions. Various details such as detailed actions, the attributes of the person including the gender, clothes and face expressions are not represented in captions.

Video-to-text (V2T), a generalized video captioning framework could be another way to approach the issue by replacing output captions with questions. Despite a video has more complete information comparing to captions as the input
, its major drawback is \textbf{the absence of the answer}. A recent work, \baseline{} \cite{srcmsa} adopted V2T to generate questions alone. Although transformer based \baseline{} outperforms various video-to-text baselines by proposing a stronger video encoder, video-question pairs do not enable video question ``answering” training. Moreover, applying V2T to QG task also suffers from factual error, repetition and redundancy problems. One of the main reasons for these problems is training with perplexity loss only. Perplexity loss is a double-edged blade for question generation as it captures question patterns but does not focus on specific details. More specifically, vital cues for question generation such as an object, an attribute, or a relation may weigh only a few or even a single word and be neglected.
\begin{table}[]
    \centering
    \begin{tabular}{c|c}
         \revision{APV} & \revision{Answer Proposal Vector}  \\
         \revision{ASV} & \revision{Answer Sheet Vector} \\
         \revision{Anet-QA} & \revision{ActivityNet-QA \cite{anetqa}}  \\
         \revision{CMSA} & \revision{Cross-Modal Self Attention} \\
         \revision{CapQG} & \revision{Caption Question Generation} \\
        \revision{GPN} & \revision{Generator-Pretester Network} \\
        \revision{JQAG} & \revision{Joint Question-Answer Generator} \\
        \revision{KLdiv} & \revision{KL Divergence} \\
        \revision{PT} & \revision{Pretester} \\
        \revision{Q.Ctrl} & \revision{Question Controller} \\
        \revision{QA} & \revision{Question Answering} \\
        \revision{SA} & \revision{Self Attention} \\
        \revision{SRCMSA} & \revision{Semantic Rich Cross-Modal Self Attention \cite{srcmsa}} \\
        \revision{TCL} & \revision{Target Consistency Loss} \\
        \revision{V2T} & \revision{Video-to-text} \\
        \revision{VQAG} &\revision{ Video Question-Answer Generation} \\
        \revision{VQG} & \revision{Visual Question Generation} \\
        \revision{VideoQA} & \revision{Video Question Answering} \\
         \hline
    \end{tabular}
    \caption{\revision{List of Abbreviations}}
    \label{tab:abbe}
\end{table}


In this work, we propose a novel yet effective \oursfull{} (\ours{}) (Figure \ref{fig:model}) which jointly generates question-answer pairs and verifies the generated question by pretesting it. 
The \ours{} model has two technical components to tackle the above challenges. 
(1) The \ourgenfull{} (\ourgen{}), which comprehends the video and provides the answer proposal according to the video and the desired question type. The answer proposal and the video are then used to generate a question-answer pair in an end-to-end manner.
(2) The \ourmodulefull{} (\ourmodule{}), which pretests the generated question and verifies the answer with the answer proposal and the answer ground truth.
Traditional V2T approaches suffer from factual errors because it optimizes the model by checking the generated question word-by-word, and encourages the model to neglect essential facts due to their low word frequency. Our \ourmodulefull{} tackles the problem in a new way by trying to answer the generated question.
Many modern human-annotated QA datasets, including text QA \cite{squad,NewsQA}, and Video QA \cite{anetqa}, perform additional manual verification after question labeling. Another set of crowd-workers is asked to answer questions according to an input video or passage. Inspired by this concept, our \ourmodulefull{} acts as an agent which pretests a generated question and checks the answer. Specifically, the \ourmodulefull{} first attempts to answer the question by projecting the question embedding to the answer space and produces an Answer Sheet Vector (ASV).  
Next, \ourlossfull{} (\ourloss{}) is performed to verify the ASV with the ground truth answer and the answer proposal.




  

We evaluate our \ours{} model on two large-scale, human annotated datasets, ActivityNet-QA \cite{anetqa} and TVQA \cite{tvqa}, and achieve the state-of-the-art performances. On ActivityNet-QA dataset, we significantly outperform previous state-of-the-art V2T and CapQG models with merely 20\% of training data. 
Our experimental results also confirm that conventional CapQG approaches suffer from information loss during machine video captioning. 
We observe a huge question generation performance drop when training with machine generated captions instead of human labeled ones, in spite of using a modern captioning model.
We also applied our generated questions to the Video Question Answering task and provide the new perspectives
of Video QA training. We reached 29.5\% of accuracy training with \textit{only VQAG produced data}, which outperforms some supervised baseline training and indicates that the knowledge obtained from generated data is able to train Video QA models. Also, we compared the fine-tuning performance with different pre-training strategies including CapQG and transfer learning. Our approach achieved the best QA performance in both semi-supervised and fully supervised scenarios.
To pave a novel path for future researchers, we carefully analyzed the quality of generated questions for Video QA training.

  
Our contributions are listed below:
    \textbf{(1)} We propose a novel task, \ourtaskfull{}, which jointly generates question-answer pairs from videos.
    %
    \textbf{(2)} We propose an end-to-end \oursfull{} which jointly generates a question-answer pair and verify the generated question by pretesting it.
    \textbf{(3)} Our proposed model reaches a new state-of-the-art question generation performance 
    on two large-scale, human annotated Video QA datasets. 
    \textbf{(4)} We apply generated question-answer pairs to Video QA applications and demonstrate the opportunity of Video QA training with VQAG generated data. 

\section{Related Work}

Video Question Answering (Video QA) has been a crucial yet challenging task in multimedia. While deep learning has enabled representation learning for multimedia features such as image, video and text and achieved remarkable performance on various tasks, one of the biggest barriers 
is the need for abundant training data. The annotation process of Video QA data requires video comprehension and question-answer labeling and is much more time-consuming compared to classification tasks such as action recognition. Most of large-scale Video QA datasets are automatically generated as the result. For example, Jang \etal{} proposed the TGIF-QA \cite{tgqa} dataset by using templates to generate 165,000 question-answer pairs for animated GIFs. Meanwhile, many of researches apply caption question answering (CapQG), which utilizes human labeled captions and text question generation systems, to produce sufficient question-answer pairs from existing datasets or available web videos. Therefore, the availability and the quality of generated questions strongly relies on captions. Zeng \etal{} proposed the Video-QA \cite{vdqa1} dataset by harvesting videos with human written descriptions from web and obtains 170,000 question-answer pairs for Video QA training with CapQG. 
Xu \etal{} extended two video captioning datasets, MSVD \cite{msvd} and MSRVTT \cite{msrvtt} with CapQG and generated MSVD-QA with 50,000 question-answer pairs and MSRVTT-QA with 240,000 question-answer pairs. Recently, several human labeled Video QA datasets were released with a lots of crowd-sourcing efforts. Lei \etal{} proposed a human annotated dataset, TVQA \cite{tvqa}, with 150,000 multi-choice question-answer pairs. Yu \etal{} extended the ActivityNet \cite{anet} dataset to open-ended Activitynet-QA \cite{anetqa} by manually labeling 58,000 question-answer pairs. \revision{Zadeh \etal{} proposed a Social-IQ \cite{rvdqa1} dataset to evaluate socially intelligent techniques with a year of human annotation period. Yi \etal{} tackled a more challenging of casual and collision reasoning with a synthetic dataset \cite{rvdqa3}}
As CapQG depends on captions while human annotation is expensive, it is necessary to automatically generate massive questions according to videos.


\revision{Apart from text generation, Some works applied question generation using visual signals. Image Question Generation (also known as visual question generation (VQG)) models \cite{vqg1,vqg2,vqg3} took an image and generate corresponding questions. While VQG models could generate questions for image QA, using VQG generated questions for Video QA lose vital temporal information. }
Video-to-text (V2T) Generation adopts captioning models to generate questions. Video captioning \cite{cap0,cap1,cap2,cap3,cap4,cap5,cap6,cap7,cap8,cap9,cap10,cap11,cap12,vcaptmm1,vcaptcvst1}, \revision{\cite{rvc1,rvc2,rvc3,rvc4,rvc5}}
is also an active research area in multimedia. 
A recent work, SRCMSA \cite{srcmsa}, adopts the video-to-text generation model to generate questions alone for Video QA training. SRCMSA proposed a transformer-based video-to-text model and demonstrates the effectiveness on TVQA \cite{tvqa} dataset. However, SRCMSA does not consider the corresponding answer; video-question pairs alone do not enable Video QA training as models cannot be optimized without answers. Moreover, V2T models trained with perplexity loss generally suffer from factual error as perplexity loss prioritizes repeating patterns rather than relatively infrequent words such as objects, attributes or relations. Hence, our proposed model tackles the above challenges by jointly generating a question-answer and verifies the question by pretesting it.


\section{\oursfull{} (\ours{})}\label{section:ours}
\begin{figure*}
  \includegraphics[width=\textwidth]{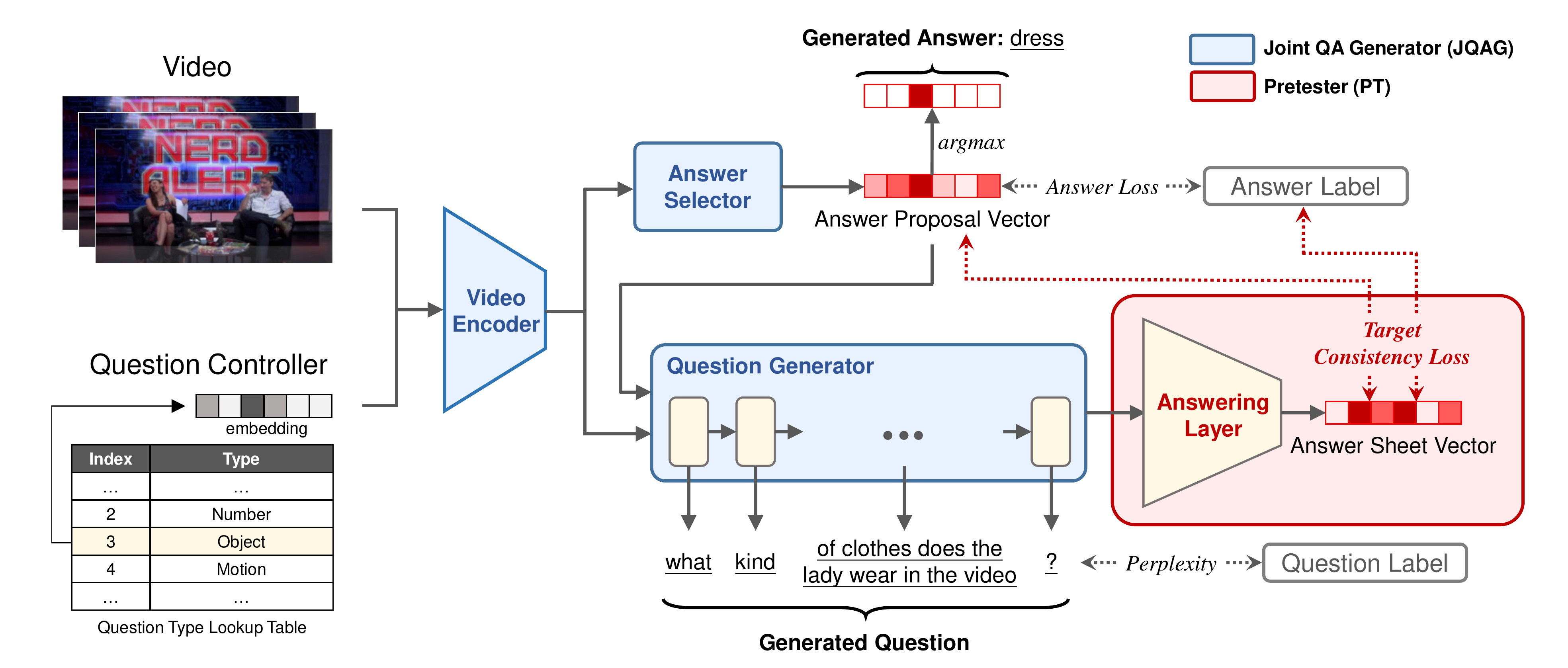}
  \caption{\oursfull{} (\ours{}) Model which generates massive question-answer pairs with two novel components. (1) \ourgenfull{} (\ourgen{}, Section \ref{subsection:JQAG}) comprehends the video and jointly generates a question-answer pair. \ourgen{} first infers an Answer Proposal vector and then generates a question-answer pair. (2) \ourmodulefull{} (\ourmodule{}, Section \ref{subsection:loss}) pretests the generated question by trying to answer it and check the answer with both the answer proposal and the answer ground truth. (cf. Section \ref{section:ours}) [Best viewed in color.]
  }
  \label{fig:model}
\end{figure*}

As shown in Figure \ref{fig:model}, our model is composed of two novel components. (1) \textcolor{blue}{Joint QA Generator} jointly generates a question-answer pair according to a video clip and the question controller by estimating an Answer Proposal Vector (APV) and feeding APV to generate a question and an answer. (2) \textcolor{red}{\ourmodulefull{}} tries to answer the generated question and checks the answer with the answer proposal and the answer ground truth. Our model takes a video and the question controller as input and generates question-answer pairs accordingly. The question controller represents a desired question type such as spatial relation and can be obtained during dataset annotation. Precisely, given a video $V$ and a question-answer type $T_{i}$, we generate a question-answer pair $Q_{i}, A_{i}$. Comparing to a previous work \cite{srcmsa} which generates \textit{a single question} based on a video clip, we are able to generate \textit{multiple question-answer pairs} for Video QA. This is essential for Video QA training as the model learns to not only comprehend but \textit{answer} the question.

The input video can be represented in various ways. 
In our work, 
we follow SRCMSA \cite{srcmsa} for fair comparison. 
The inputs of our model are following features extracted from $n$ frames: (1) CNN extracted visual feature \revision{${V}^{F}=\{{V}^{F}_{1}, {V}^{F}_{2}, \cdots , 
{V}^{F}_{n}\}$, $V^{F}_{i} \in{\mathbb{R}}^{2048}$}, (2) Object features 
\revision{${V}^{O}=\{{V}^{O}_{1}, {V}^{O}_{2}, \dots , {V}^{O}_{n}\}$}, where $V^{O}_{i} \in{\mathbb{R}}^{256}$, and (3) Question Controller $C\in{\mathbb{R}}^{256}$ which determines the type of generated questions. The model outputs (1) Question \revision{${Q}=\{{Q}_{1}, {Q}_{2}, \dots , {Q}_{m}\}$}, ${Q}_{i} \in{\mathbb{R}}^{|V|}$, where $m$ is the word length of the questions and $|V|$ is vocabulary size, and \revision{(2) Answer $A \in R^{|A|}$} where $|A|$ represents answer size.  
In our experiments, we extract 20 video frames per video following existing Video QA settings \cite{vdqa0, anetqa}. Then, we obtain CNN features from the Pool5 layer of ResNet101 \cite{resnet} and detect objects features by Faster-RCNN \cite{DBLP:journals/pami/RenHG017} pretrained on Visual Genome \cite{visualgenome}. Afterward, we mean pool the object representation in each frame and get object embeddings.


The model optimizes the trainable parameters as follow:
\begin{equation}
Q, A = f(\theta_{E}, \theta_{QG}, \theta_{A}, \theta_{AS}, {V}^{F}, {V}^{O}, C),
\end{equation}
where $\theta_{E}, \theta_{QG}, \theta_{A}, \theta_{AS}$ are the parameters of Video Encoder, Question Generator, Answering Layer and Answer Selector.

\subsection{Joint Question-Answer Generator (JQAG)}\label{subsection:JQAG}
\subsubsection{Video Encoder}
The Video Encoder encodes video features and the question controller to generate a question-answer pair.
We adopt a transformer-based Cross-Modal Self-Attention (CMSA) Encoder \cite{srcmsa}. 
The video encoder first fuses frame feature $V^{F}$ and object feature $V^{O}$ with a projection $W_{proj} \in{\mathbb{R}}^{2048 \times 256}$ for each frame:
\begin{equation}
    \revision{V^{S}_{i} = W_{proj}(V^{F}_{i}) \odot V^{O}_{i}},
\end{equation}
where $V^{S}_{i} \in{\mathbb{R}}^{256}$.

The original CMSA encoder does not consider the question type. However, different types of questions require different signals in the video. For instance, a question about spatial relationship relies on certain objects, while the question about the scene needs the cues of whole frame. Observing this, we introduce the Question Controller $C$ which represent the question type with a 256 dimension embedding for each frame:

\begin{equation}
    \revision{V^{src}_{i} = V^{S}_{i} \odot C}
\end{equation}

\revision{C is obtained from a question controller matrix $C^{M}\in{\mathbb{R}}^{256 \times T}$, where $T$ represents the number of question types.}

$V^{src}$ is then fed into Self-Attention Layers:
\begin{equation}
    \revision{V^{i+1} = SA(W^{Q}_{i}V^{i}, W^{K}_{i}V^{i}, W^{V}_{i}V^{i}),}
\end{equation}
\revision{where $V^{i} \in{\mathbb{R}}^{256 \times n}$ is the $i$-th layer output, $SA$ refers to multi-head self attention. We initialize the first layer by $V^{0} = PE(V^{src})$ and $PE$ is positional encoding introduced in previous work \cite{transformer}.}

\revision{Finally, we mean-pool the last layer output to obtain the clip level representation, and feed the clip level representation into a two-layer projection:}
\begin{equation}
    \revision{V^{final} = W^{P}_{2}(ReLU(W^{P}_{1}meanpool(V^{|L|})))}
\end{equation}
to obtain the video embedding for answer and question generation.

\subsubsection{Answer Selector} \label{subsection:AG}
The Answer selector aims to produce the answer proposal vector according to the video. Video QG with answer allows Video QA training based on generated question-answer pairs. The answer selector takes video embedding $V^{final}$ and feeds forward the networks includes two trainable linear layers $W^AG_{1}$ and $W^AG_{2}$:
\begin{equation}
    A^{P} = softmax(W^{AG}_{2}(ReLU(W^{AG}_{1}V^{final})))
\end{equation}

The answer proposal vector $A^{P}$ represents the distribution of the answer likelihood for each candidate answer. During inference stage, the answer is generated by applying $A = argmax(A^{P})$. \revision{During training stage, $A^{P}$ is optimized by Cross Entropy Loss:}
\begin{equation}
    \revision{L^{ap} = CrossEntropy(A^{P}, A^{tgt}),}
\end{equation}
\revision{where $A^{tgt}$ denotes the ground truth answer.}

\subsubsection{Question Generator}
The Question Generator takes the video embedding $V^{final}$ and the answer proposal $A^{P}$, and outputs the question according to them.
We utilize the LSTM decoder including two hidden layers and an output layer used in previous work \cite{srcmsa}. \revision{In addition, for the Video QG task, question composition is less complex than the input video, so we reduce the parameters. For example, Anet questions are mostly less than 10 words on average. Also, TVQA questions, while being longer, follow a strict pattern of ``Something before/after/then Something." Hence, compared to general video-to-text tasks, question generation follows question patterns and therefore is much simpler.}

\revision{The question is generated word-by-word according to the previous hidden state:}
\begin{equation}
    Q_{i+1}, H_{i+1} = LSTM(Q_{i}, H_{i}),
\end{equation}
where $Q_{i}$ is the \revision{$i$-th} word and $H_{i}$ is the \revision{$i$-th} hidden state.
We initialize the hidden states in LSTM with video embedding and the answer proposal:
\begin{equation}
    H_{0} = V^{final} \odot W^{AE}(A^{P}),
\end{equation}
where $H_{0}$ is the initial hidden states and  $W^{AE}\in{\mathbb{R}}^{|A| \times 256}$ embeds the generated answer proposal. This constrains the question generator with the input video and the model proposed answer. 
\revision{The loss of question generator is obtained by the perplexity of generated question $Q$ and the ground truth question $Q^{tgt}$:}
\begin{equation}
    \revision{L^{qg} = Perplexity(Q, Q^{tgt})}
\end{equation}
Finally, we keep last hidden state $H_{|Q|}$ as question embedding $Q^{emb}$.

\subsection{\ourmodulefull{} (\ourmodule{})} \label{subsection:loss}
\ourmodulefull{} (\ourmodule{}) pretests the generated question $Q^{emb}$ and generates an Answer Sheet Vector $A^{Q}$, and then optimizes the consistency between $A^{Q}$, the answer proposal $A^{P}$, and the ground truth answer $A^{tgt}$. First, the Answering Layer acts as an agent to answer the question and generates the Answer Sheet Vector $A^{Q}$:
\begin{equation}
    A^{Q} = softmax(W^{AL}_{2}(ReLU(W^{AL}_{1}Q^{emb})))
\end{equation}

Then, we apply our \ourlossfull{} (\ourloss{}). $A^{Q}$ is optimized by (1) the ground truth answer with Cross Entropy Loss:
\begin{equation}
    L^{ans} = CrossEntropy(A^{Q}, A^{tgt})
\end{equation}

and (2) the Answer Proposal Vector by minimizing KL-divergence (KLdiv):
\begin{equation}
    L^{c} = KLdiv(A^{Q}, A^{P})
\end{equation}

\revision{Finally, the \ourlossfull{} (TCL) is obtained by
\begin{equation}
    L^{tc} = \lambda_{c}L^{c} + \lambda_{a}L^{ans},
\end{equation}
where $\lambda_{c}$ and $\lambda_{a}$ are hyper-parameters and $\lambda_{c}$ + $\lambda_{a}$ = 1.}

\revision{The final loss is the sum of the TCL, the perplexity loss and the answer loss:
\begin{equation}
    L^{total} = L^{tc}+L^{qg}+L^{ap}
\end{equation}
}




\section{Question Generation Experiments} \label{section:qg}
In this section, we adopt experiments on Video Question Generation and analyze the results in terms of quantity and quality. The results demonstrate the effectiveness of our proposed model. Furthermore, we carefully analyze the generated questions and lead a path for VQAG researches. The code is available at https://github.com/htsucml/VQAG

\subsection{Experimental Setup}
\subsubsection{Data} \label{subsubsection:data}
\begin{table} 
    \centering
    \begin{tabular}{
    |c|c|
    c|c|c|c|
    }
    \hline
    
    & QG-train 
    & QG-val
    & QG-test
    & QA-val
    & QA-test
    \\ \hline
    Anet-QA
    & 14,400 
    & 2,880
    & 14,400
    & 18,000
    & 8,000
    \\ \hline
    TVQA
    & 122,093 
    & 15,253
    & 7,623
    & -
    & -
    \\ \hline
    \end{tabular}
    \caption{Data split. For Activitynet-QA (Anet-QA) \cite{anetqa}, we split the original training set into QG-train/QG-val/QG-test in the ratio of 45\%/10\%/45\% for QG experiments in Section \ref{section:qg}. QG-test set is also used for pre-training of our QA experiments in Section \ref{section:qa}. For TVQA, we follow the setting of \cite{srcmsa}.
    }
    \label{table:datastat}
\end{table}

We use two recent human annotated datasets, including ActivityNet-QA (Anet-QA) \cite{anetqa} and TVQA \cite{tvqa} for Video QG experiments. The statistics of the dataset is shown in Table \ref{table:datastat}. The main dataset for Video QG and QA experiments is Anet-QA, which is an open-ended dataset, as TVQA dataset is a multi-choice dataset which involves distractors.  
We also evaluate Video QG results on TVQA dataset.

%
To evaluate both Video Question Generation and Video Question Answering (Section \ref{section:qa}), we split the original Anet-QA training set into \textit{QG train}, \textit{QG val} and \textit{QG test} sets. For TVQA dataset, we follow the setting of a previous work \cite{srcmsa}. 

\subsubsection{Evaluation Metrics}
Following previous approaches, we evaluate Video QG with BLEU \cite{BLEU} (B), BLEU-4 (B4), ROUGE \cite{ROUGE} (R), CIDEr \cite{CIDEr} (C) and METEOR \cite{meteor} (M) scores. BLEU is a word-level precision metric, and BLEU-4 is the BLEU score for 4-grams. ROUGE is a word-level recall measurement. CIDEr considers word frequency and penalizes frequent words by TF-IDF weighting. METEOR takes synonyms into consideration by utilizing WordNet \cite{wordnet}.

\subsubsection{Implementation Detail}
We implement our model and the baseline with OpenNMT \cite{OpenNMT} built on Pytorch \cite{paszke2017automatic}. \revision{For fair comparison, we add the question controller to all baselines. 
For V2TQG, we modify the CMSA encoder (see Section \ref{subsection:JQAG}) and add the question controller, which follows the same approach of our method. For CapQG and CapQG-oracle, we add all question types as tokens in the vocabulary and concatenate the question with the corresponding question type token. We set $\lambda_{c}=0.25$ and $\lambda_{c}=0.75$ for TCL. We use ADAM \cite{adam} as our optimizer and set $\alpha=0.01$, $\beta1=0.9$, and $\beta2=0.999$. The learning rate is initialized as $1e-3$ for Anet-QA and $2e-5$ for TVQA. We train the model for 200,000 steps for fully supervised experiments, and train the model for 40,000 steps for semi-supervised (20\%) experiments. The model is validated every 5,000 steps. For fair comparison, we use the same hyper-parameters for the V2TQG baseline. We use the default parameters for CapQG and CapQG-oracle baselines.}

\subsection{Question Generation Performance}

\subsubsection{Activitynet-QA Results}
\label{subsection:results:anet}
\begin{table} 
    \centering
    \begin{tabular}{
    |l|c|
    c|c|c|c|c|
    }
    \hline
    %
    & B
    & B4
    & R
    & C
    & M
    \\ \hline
    %
    CapQG 
    (100\%)
    &  67.63 & 40.64 & 62.39 & 8.56 & 29.75
    \\
    CapQG-oracle* 
    (100\%)
    &  71.17 & 44.58 & 63.76 & \textbf{9.49} & 31.28
    \\
     %
    V2TQG 
    (100\%)
    &  67.95 & 40.64 & 61.95 & 7.21 & 30.41
    \\
    \hline
    \ours{} (ours) (20\%)
    & 69.49 & 42.81 & 63.19 & 7.68 & 31.31
    \\
    \textbf{\ours{} (ours)} \textbf{(100\%)}
    & \textbf{72.38} & \textbf{46.00} & \textbf{66.22} & 9.32 & \textbf{32.39}
    %
    \\
    %
%
    \hline
    \end{tabular}
    \caption{QG Results on Activitynet-QA dataset. Our model outperforms CapQG and V2TQG baselines. B:BLEU, B4:BLEU-4, R:ROUGE, C:CIDEr, M:METEOR. *CapQG-oracle is a very strong baseline which uses ground truth captions, which are not always available practically.
    See Section \ref{subsection:results:anet} for detailed discussion.
    }
    \label{table:qg}
\end{table}

Table \ref{table:qg} examines the Question Generation performance on Activitynet-QA \cite{anetqa}. We compare our method with both Caption QG (CapQG) and video-to-text QG (V2TQG). For CapQG, we generate captions with a competitive dense captioning model, Masked Transformer \cite{densecap}, and then apply the state-of-the-art text QG model, UniLM \cite{unilm} for question generation with generated captions and ground truth answers. We sort generated captions based on start time and remove duplicated captions. We also compare with the oracle case where captions are human annotated (CapQG-oracle). For V2TQG, we compare with the modern SRCMSA \cite{srcmsa} model.

Our model reaches the new state-of-the-art and outperforms all competitive baseline in a large margin as shown in the second block, including a very strong CapQG-oracle which uses human labeled captions and a BERT \cite{bert} based question generation model. Our full model training with merely 20\% of data (second row) significantly surpasses both CapQG and V2TQG training with 100\% of data in all metrics but CIDEr. As \ours{} does not require abundant data to work well, it is applicable in practice where the annotated data is available but expensive.

Comparing baseline models CapQG and CapQG-oracle in the first block in Table \ref{table:qg}, we could see more than 3 points of BLEU scores, roughly 1.5 points of ROUGE and METEOR, and about 1 point of CIDEr drop when using generated captions, even with a modern captioning system. It demonstrates that CapQG suffers information loss in practice where human labeled text descriptions are unavailable.  

\subsubsection{TVQA Results} \label{subsection:results:tvqa}
\begin{table} 
    \centering
    \begin{tabular}{
    |l|c|
     c|c|c|c|
    }
    \hline

    & B
    & B4
    & R
    & C
    & M
    \\ \hline
    S2VT$\dagger$ \cite{S2VT} 
    &  57.80 & 7.58 & 36.25 & 6.39 & 14.83
    \\
    OBJ2TEXT$\dagger$ \cite{obj2text} 
    &  61.78 & 10.44 & 38.49 & 6.42 & 15.33
    \\
    IMGD$\dagger$ \cite{imgd} 
    &  61.08 & 9.59 & 37.78 & 7.29 & 15.21
    \\
    \baseline{} \cite{srcmsa} 
    & 66.11  & 12.17 & 42.02 & 28.81 & 18.82
    \\
    \hline
%
    \textbf{\ours{} (ours)}
    & \textbf{68.14}  & \textbf{13.63}  & \textbf{42.95}  & \textbf{30.87}  & \textbf{19.53} 
    \\
    \hline
    \end{tabular}
    \caption{Video QG Results on TVQA. $\dagger$: obtained from \cite{srcmsa}. For the fair comparison, we implement SRCMSA and run the experiments following their setting with same random seed. Our model significantly outperforms all baselines, including previous state-of-the-art in a large margin. See Section \ref{subsection:results:tvqa} for detailed discussion.} 
    \label{table:tvqa}
\end{table}



Table \ref{table:tvqa} shows performance comparison with S2VT \cite{S2VT}, IMGD \cite{imgd}, OBJ2TEXT \cite{obj2text}, and \baseline{} \cite{srcmsa} on TVQA \cite{tvqa} dataset. \ours{} outperforms state-of-the-art results across all metrics and attains 19.5 of METEOR score. Different from ActivityNet-QA, TVQA dataset includes both video and subtitles input. The questions are also longer comparing to ActivityNet-QA. The remarkable results demonstrate the robustness of our proposed method.

\subsection{Ablation Study}\label{subsection:ablation}
The ablation study is shown in Table \ref{table:qg_ana}. Removing \ourmodulefull{} module (Section \ref{subsection:loss}) results in performance drop in all metrics especially CIDEr and METEOR scores. Comparing BLEU and ROUGE scores, which match word level precision and recall, CIDEr considers word frequency and penalizes repeating words, and METEOR takes synonyms into account. Hence, it reveals that \ourmodulefull{} boosts the performance by prioritizing the essential cues to answer more than simply matching words and observed patterns. 

\revision{The question controller (Q.Ctrl) improves the performance in all metrics except for ROUGE, especially for more than 2\% of METEOR score. Without the question controller (w/o Q.Ctrl), the video QAG model learns one-to-many mapping with multiple questions associated with a video, resulting in higher matching scores on frequent patterns at the price of rare patterns and diversity. Therefore, the w/o Q.Ctrl model obtains comparable BLEU, ROUGE, and CIDEr scores while the performance drops when considering synonyms, such as METEOR score. On the other hand, with the aid of the question controller, the Video QAG model needs to perform one-to-one mapping and is forced to capture the semantics instead of repetitive patterns, resulting in a more significant METEOR score gap. In addition, Question Controller enables the model to control the desired question type during the inference stage.
}
\begin{table} 
    \centering
    \begin{tabular}{
    |l|ccccc|c|
    }
    \hline
    
    & B
    & B4
    & R
    & C
    & M
    & $\triangledown$M
    \\ \hline
    Full,  
    $\lambda_{a}=0.75$
    &  \textbf{69.49}  &  \textbf{42.81} &  63.19 &  \textbf{\revision{7.68}} &  \textbf{31.31} &
    -
    \\
    \hline
    No \ourmodulefull{}
    & 69.16  & 42.25  & 63.03  & 7.05  & 30.94 & 1.18\%
    \\
    No Object
    & 69.15  & 42.31 & 63.03  & 7.05  & 30.94 & 1.18\%
    \\
    No Frame
    & 69.06  & 42.29 & 63.11 & 6.73 & 31.02 & 0.93\%
    \\
    \revision{No Q.Ctrl}
    & \revision{69.38} & \revision{42.73} & \textbf{\revision{63.33}} & \revision{7.65} & \revision{30.52} & \revision{2.52\%}
    \\
    \hline
 $\lambda_{a}=0.00$
    & 68.60 & 41.39 & 62.20 & 7.05 & 30.44 & 2.77\%
    \\
 $\lambda_{a}=0.25$
    & 68.97  & 41.69 & 62.96 & 7.03 & 31.09 & 0.70\%
    \\
 $\lambda_{a}=0.50$
    & 69.47  & 42.57 & 63.13 & 7.12 & 31.26 & 0.16\%
    \\
 $\lambda_{a}=1.00$
    & 68.84  & 41.82 & 62.96 & 7.00 & 31.15 & 0.51\%
    \\
    \hline
    \end{tabular}
    \caption{\revision{Q.Ctrl: Question Controller} Ablation Study with 20\% of ActivityNet-QA QG-train set used, comparing with full model and best hyper-parameters (First block). Second block examines input features and components and shows all of them contribute to the performance. Third block tests different hyper-parameters of \ourmodulefull{} module (Section \ref{subsection:loss}) and demonstrates the best performance using both Answer Proposal and Answer Ground Truth as labels. (cf. Section \ref{subsection:ablation}). 
    }
    \label{table:qg_ana}
\end{table}

Removing video or object features also leads to performance drop as both video frames and objects provide cues for \ourtask{}. Video frames provide complete but raw and sparse visual cues, while objects allow \ourtask{} model to capture word semantics according to object co-occurrence. Some of the questions, such as scene (day/night) related question-answer pairs 
require frame features to generate. Meanwhile, some questions rely on the semantics of only certain objects.  
Questions involve spatial 
or temporal relationships 
entails both video frames and visual concepts to properly pinpoint the essential object semantics and comprehend the relationship between them.

The third block of Table \ref{table:qg_ana} analyzes the performance depending on the parameters $\lambda_{A}$ and $\lambda_{C}$ for \ourloss{} in the \ourmodulefull{} module (Section \ref{subsection:loss}), which controls the loss obtained from ground truth answer and model generated answer distribution. We reach best performance when setting $\lambda_{A} = 0.75$ and $\lambda_{C} = 0.25$. Furthermore, we can observe that performance drops when either loss is removed, revealing that both signals are essential and collaborates well.

\subsection{Case Study} \label{subsection:results:quality}
\begin{figure*}
  \includegraphics[width=\textwidth]{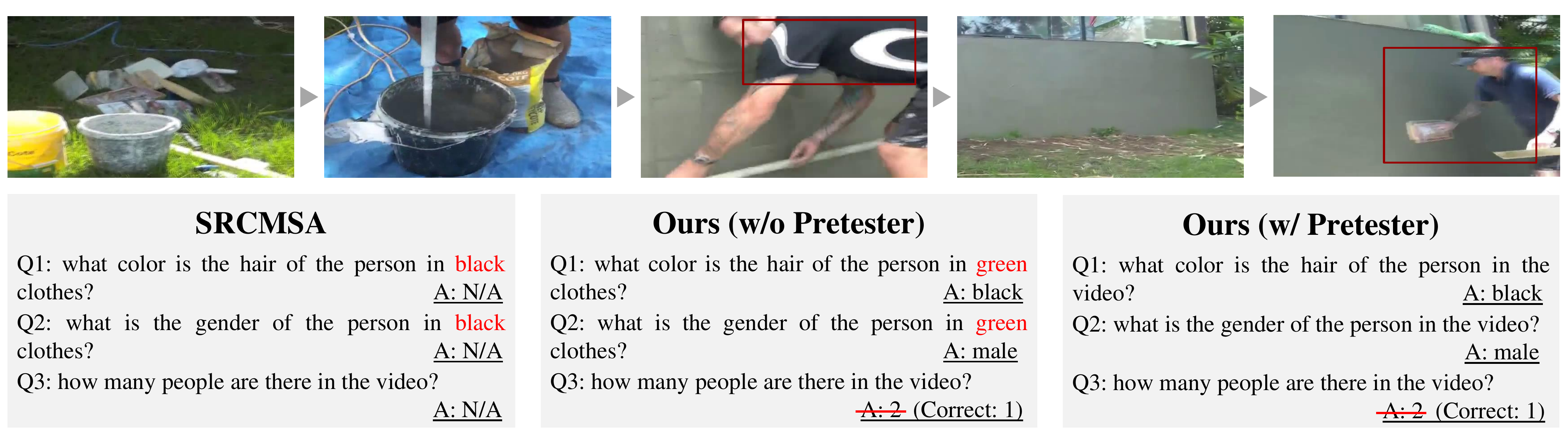}
  \caption{Case study. SRCMSA \cite{srcmsa}, a V2TQG model, only generates questions alone. Our model without the \ourmodulefull{} module (Middle) can generate question-answer pairs, but fail to capture correct details (the color of clothes). Our full model with \ourmodulefull{} (Right) generates correct questions for Q1 and Q2. For Q3, both variants of our model failed to generate the correct answer, it may be caused by the person wearing different color of clothes and frequently camera rolling in the clip. (cf.Section \ref{subsection:results:quality}) [Best viewed in color.] 
  }
  \label{fig:cases}
\end{figure*}
\begin{figure*}
  \includegraphics[width=
  \textwidth
  ]{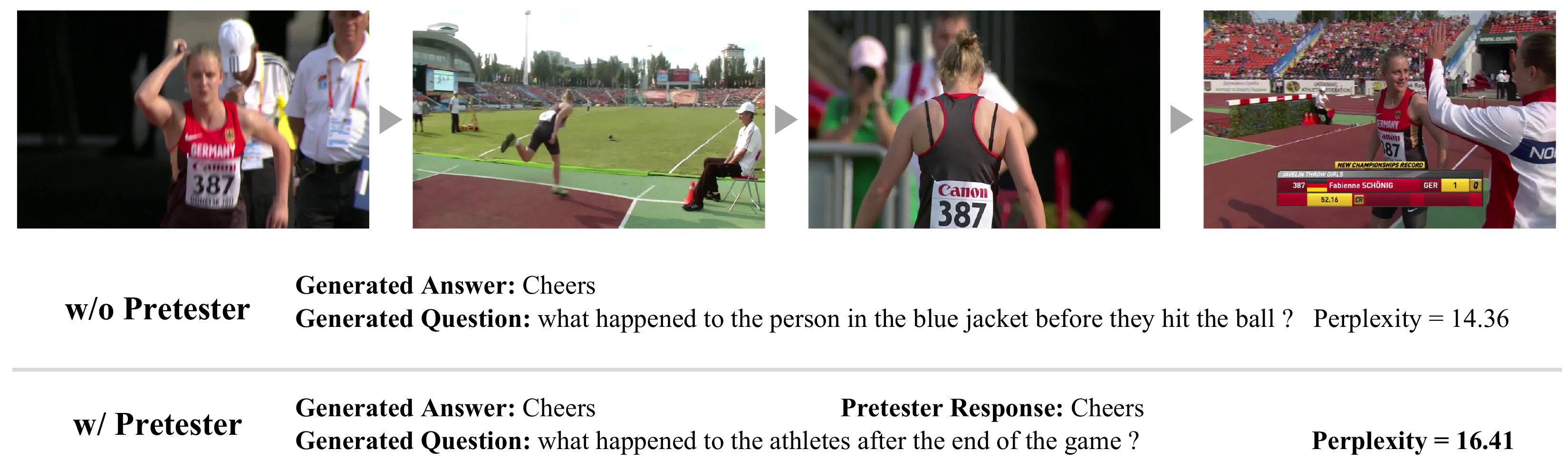}
  \caption{
  (Top) With the \ourmodulefull{} removed, the model generates a wrong question. 
  (Bottom) With the \ourmodulefull{} equipped, it encourages the generator to output a correct question in spite of higher perplexity loss. (cf. Section \ref{subsection:results:quality})
  }
\label{fig:modelana}
\end{figure*}
Figure \ref{fig:cases} illustrates several generated questions of \baseline{} (Left), our model without the \ourmodulefull{} module (Middle), and our full model with \ourmodulefull{} (Right). For question (1) and (2), while \baseline{} is able to generate questions according to the video, the absence of answers is the biggest gap to Video QA training. Additionally, \baseline{} generates questions with redundant relation ``black”. While questions are literally not wrong, redundant words may distract Video QA models to pay attention on useless details. Middle block exhibits the results of our model with \ourmodulefull{} removed. It still successfully outputs the answer, but with an incorrect relation in each question. Without the \ourmodulefull{}, QAG model fails to focus on a key object (the person in this case) and is distracted by grass (colored green) in the video, and generates wrong questions. On the right block, our model guided with \ourmodulefull{} generates precise questions without redundancy. For question (3), each model successfully generates a question, while both variants of our models output a wrong answer 2 (should be 1). It may cause by the only man wearing different color of clothes, and frequent lens rotation when he switches inside and outside the view. Above results indicate that while we significantly outperform previous state-of-the-art models, Video QAG remains challenging due to the sparsity, the diversity, and the temporal dependencies of video features.

Figure \ref{fig:modelana} demonstrates generated questions and perplexity scores. We compare our models with or without \ourmodulefull{}. Our model without \ourmodulefull{} generates an unanswerable, wrong question with non-existent facts (in the blue jacket, hit the ball). Our full \ours{} equipped with \ourmodulefull{} generates an answerable question despite a higher perplexity. 
It reveals that the \ourmodulefull{} encourages the generator to generate an answerable question by ignoring frequent but wrong words during decoding.

\subsection{Error Analysis} \label{subsection:results:error}
\begin{figure*}
  \includegraphics[width=\textwidth]{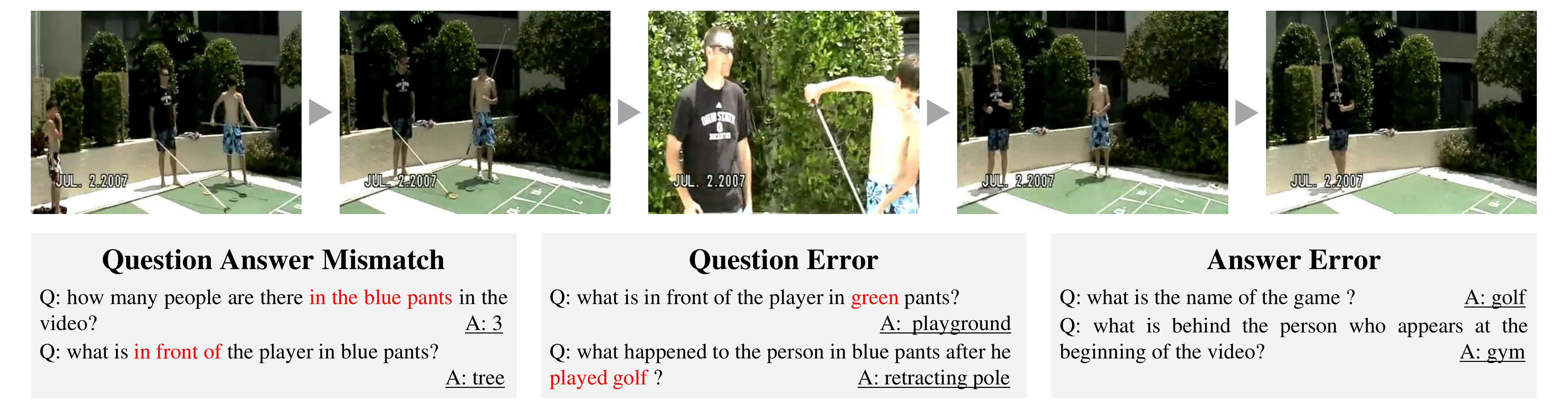}
  \caption{Error Analysis. We dig into the generated questions and analyze the occurred errors. Left: QA Mismatch which the question and the answer both are in the video clip but does not match. 
  Middle: Question Error where the question is invalid according to the video. Right: Answer Error where the answer is invalid according to the video. (cf. Section \ref{subsection:results:error}) [Best viewed in color.]
  }
\label{fig:errors}
\end{figure*}
We achieve state-of-the-art performance for the question generation task. Additionally, we investigate errors of generated questions and classify errors into 3 categories. \textit{QA Mismatch Error, Question Error and Answer Error}, as shown in Figure \ref{fig:errors}. 

\textbf{Question-Answer (QA) mismatch error} (Left block) is where a question and an answer are both represented in the video but do not match properly. For instance, an answer of first question-answer should be ``1” instead of ``3” as there is only a person in blue pants. Otherwise, we have to change this question to ``how many people are there in the video” to match the answer ``3”. Similarly, for the second one, we can either adjust the answer to ``court”, or modify the question by replacing ``in front of” with ``behind”. QA Mismatch Error indicates that the VQAG model successfully comprehends a video clip and capture essential segments but fails to pair them. QA pairs with this type of error are usually seen in videos with various objects. This suggests that \ourmodulefull{} is a promising direction that worth considering as it directly measures the question by trying to answer it.

\textbf{Question Error} (Middle block) represents invalid questions which are undesirable on the basis of a video clip despite the corresponding answer. Previous work \cite{srcmsa} has mentioned some types of question errors, including unanswerable, redundant, and general questions. The most severe type are unanswerable questions as they may mislead the video QA model. We observed that many question errors are mostly caused by a single word of the wrong attribute or action. In other words, these unanswerable questions are actually ``mostly correct” in terms of automatic evaluation such as BLEU or ROUGE scores. This suggests future research should consider weighted or semantic metrics such as CIDEr and METEOR scores, and carefully analyze qualities beyond scores.

\textbf{Answer Error} (Right Block) is defined as an invalid answer according to a video clip in spite of a question. It is worth mentioning that most of Answer Errors occurred together with Question Errors, or pair with more general questions such as ``what is the name of the game ?”. The phenomenon reveals that Answer Errors are usually presented when the model failed to comprehend a video clip. In this situation, a question is either wrong or ``safely guessed” as frequent and general questions by the distribution learned by language modeling. This reveals that the model fails to understand the input video and suggests more fine-grained comprehension of the video encoder.

\section{Application in Question Answering} \label{section:qa}
\begin{table*}
    \centering
    \begin{tabular}
    {|llc|c|c|c|c|c|c|c|}
    \hline
    &   &
    & Motion
    & Spatial
    & Temporal
    & Y/n
    & Counting
    & Free
    & All
    \\ \hline
    \multirow{3}{*}{Unsupervised} & Random 
    & 
    & 0.1 & 0.1 & 0.1 & 0.1 & 0.1 & 0.1 & 0.1
    \\ 
    & Question Type Prior
    & 
    & 0.5 & 0.5 & 0.5 & 50.0 & 10.0 & 2.6 & 18.7
    \\ 
    & \textbf{\ours{} (ours) + HME}
    &
    & 2.5 & 7.0 & 0.9 & 50.5 & 45.6 & 20.1 & \textbf{29.5}
    \\ \hline
    \multirow{3}{*}{20\% Supervised}
    & HME \cite{vdqa5}
    & 
    & 2.5 & 5.6 & 3.5 & 60.2 & 42.0 & 41.0 & 33.3
    \\ 
    & MSVD-trans + HME \cite{vdqa5} 
    &
    & 8.5 & 8.1 & 3.1 & 57.3 & 40.8 & 40.2 & 33.4
    \\
    & CapQG-oracle \cite{unilm} + HME \cite{vdqa5}
    &
    & 4.8 & 6.6 & 2.6 & 58.5 & 39.8 & 25.2 & 33.6
    \\
    & \textbf{\ours{} (ours) + HME \cite{vdqa5} }
    &
    & 4.8 & 6.6 & 2.6 & 61.3 & 34.5 & 41.9 & \textbf{34.3}
    \\ \hline
    \multirow{6}{*}{100\% Supervised}
    & E-VQA$\dagger$ \cite{VQA}
    & 
    & 2.5 & 6.6 & 1.4 & - & - & 34.4 & 25.1
    \\ 
    & E-MN$\dagger$ \cite{EMN}
    & 
    & 3.0 & 8.1 & 1.6 & - & - & 36.9 & 27.1
    \\ 
    & E-SA$\dagger$ \cite{ESA}
    & 
    & 12.5 & 14.4 & 2.5 & - & - & 41.2 & 31.8
    \\ 
    & HME \cite{vdqa5}
    & 
    & 14.3 & 12.2 & 7.0 & 62.7 & 44.1 & 46.6 & 39.4 
    \\ 
    & MSVD-trans + HME \cite{vdqa5}
    & 
    & 16.1 & 13.5 & 7.9 & 60.6 & 38.5 & 44.7 & 38.2
    \\
    & CapQG-oracle \cite{unilm} + HME \cite{vdqa5}
    & 
    & 7.5 & 10.2 & 7.0 & 63.0 & 43.8 & 29.4 & 37.7
    \\
    & \textbf{\ours{} (ours) + HME \cite{vdqa5}}
    & 
    & 11.8 & 15.6 & 8.7 & 61.5 & 45.3 & 47.2 & \textbf{40.1}
    \\
    \hline
    \end{tabular}
    \caption{
    Video QA accuracy on Activitynet-QA QA-test set with QG generated data. $\dagger$: obtained from \cite{anetqa} and training with full training set (about twice of QA-train set, see Table \ref{table:datastat}).
    Training HME model with our generated data reaches remarkable 29.5\% accuracy, which even outperforms E-VQA and E-MN models training with full training set without witnessing a single human labeled pair. In 20\% supervised and 100\% supervised scenario, we also outperform all pretraining baselines.
    (cf. Section \ref{subsection:qascratch})
    }
    \label{table:qa}
\end{table*}    
    

In this section, we apply our generated question to downstream open-ended Video QA task. We demonstrate remarkable performance gain with generated questions only, and display improvement on both semi and fully supervised scenarios, and point out promising directions for Video QA training.

\subsection{Setup}
For open ended Video QA experiments, we evaluate with the accuracy in each question type as defined by \cite{anetqa}. We separate Y/N and counting questions as the answer spaces for these two types are much smaller. To simulate a practical scenario where training data is expensive but available, 
We train CapQG or VQAG models with 20\% of QA-train set.
We use a modern Video QA model HME \cite{vdqa5}.
The word embedding layer for a question is initialize by 300 dimension GLOVE \cite{glove} vectors. On pre-training stage, we use QA train (see Table \ref{table:datastat}) set. When fine-tuning, we use QG-train set as those QA pairs are resources available for both QG and QA training. For every experiment, all layers except for the final classification layer are initialized with pre-trained weights, if a word embedding is not found in the pre-trained checkpoint, we initialize that word with 300 dimension GLOVE \cite{glove}. 

\subsection{Question Answering Performance} \label{subsection:qascratch}
Table \ref{table:qa} shows Video QA accuracy on the Activitynet-QA dataset with baselines including E-VQA \cite{VQA}, E-MN \cite{EMN}, E-SA \cite{ESA} and HME \cite{vdqa5}. We also apply different pre-training strategy: (1) pre-training with MSVD-QA \cite{vdqa0} dataset (MSVD-trans) and (2) pre-training with CapQG-oracle generated questions (CapQG-oracle). 

As shown in the first block, without observing any human annotated pairs, the model trained with our generated data can achieve 29.5\% accuracy, which even outperform E-VQA and E-MN training with tens of thousands of human labeled QA pairs. 
Notably, in spatial and counting questions, we even outperform supervised training accuracy. However, our model marginally improves on Y/N questions, which may result from the less information provided from the answer during Video QG training. To improve Y/N question generation, we suggest to integrate our model with reasoning model in the future.


The second and the third block of Table \ref{table:qa} represents the fine-tuning results with 20\% and 100\% QA-train set. Our pre-training approach outperforms both MSVD-transfer and CapQG-oracle, especially when fine-tuning with full QA-train set where MSVD-transfer and CapQG-oracle degrade the performance. This indicates that generated questions enhance QA performance despite containing noise compared to human annotated ones. The performance drop of MSVD-trans with 100\% supervised reveals the domain gap of video matters on Video QA tasks. Meanwhile, CapQG-oracle suffers 17\% of accuracy drop on free type questions, which may involve details that do not cover in text descriptions. Compared with two pre-training strategies, our approach directly generate questions with videos in the same domain and minimize information loss and domain gap.


\subsection{\revision{Pre-training Data Analysis}}\label{subsection:ptdata}
\begin{table}[]
    \centering
    \begin{tabular}{|c|c|ccccc|}
    \hline \revisionrow{}
       VDQG Data  & No Pretrain & 20\% & 40\% & 60\% & 80\% & 100\%  \\ 
      w/o ft   & 0.1 & 5.7 & 11.6 & 25.0 & 27.1 & 29.5 \\
      w/ 20\% ft & 33.3 & 36.0 & 36.3 & 34.4 & 34.1 & 34.3 \\
      \hline
    \end{tabular}
    \caption{\revision{Anet-QA performance with different VDQG generated question scale using HME \cite{vdqa5}. ft: fine-tuning with annotated data. Our generated questions constantly enhances QA performance. cf. Section \ref{subsection:ptdata}}}
    \label{tab:qgdata}
\end{table}

\revision{Table \ref{tab:qgdata} presents the impact of generated questions in different scales. We evaluate different scales of VQAG generated questions in two scenarios: (1) w/o fine-tuning (training QA with QG generated data only, total 14,400 examples) and (2) w/ 20\% fine-tuning (pre-training QA with QG generated data, then fine-tuning with 20\% of human-annotated data). 
As shown in the table, pre-training with our generated questions constantly enhances Video QA performance regardless the scale. Without fine-tuning on human annotated data, QA performance is roughly correlated with the questions' scale before reaching 60\%, demonstrating that the Video QA system could benefit from different patterns from more generated questions. With fine-tuning, Video QA performance improves the most at 40\% (with 3.0 QA accuracy improved). With more generated questions adopted, we get roughly 1.0 QA accuracy gain,  revealing the gap between QG generated data and human-annotated data, and the noise in generated data might influence the QA model. Note that in spite of potential influence of the noise, pre-training with our generated questions continuously improves the Video QA performance (from 0.8 to 3.0). We suggest future research to tackle the challenge to enable more benefit from generated questions.
}

\section{Conclusion}
In this paper, we introduce a novel task,  \ourtaskfull{}, to automatically generate question-answer pairs for Video Question Answering training. We propose a novel \oursfull{} to end-to-end jointly generates question-answer pairs and verify the generated question by trying to answer it. 
We demonstrate the efficacy of proposed modules in the Video Question Generation task and achieve state-of-the-art performance on two large-scale, human annotated datasets. 
The extensive results on Video QA exhibits the performance boost from our generated questions, reaching 29\% of accuracy without witnessing any annotated data and also enhances the performance in both semi and fully supervised scenarios. To lead future research, we also provide detailed analysis of generation errors and its impact on Video QA tasks. 
\revision{We encourage future work to extend our research based on our results and analysis. For example, to resolve unanswerable questions discussed in Section \ref{subsection:results:error}, one might want to obtain more fine-grained object-level cues instead of applying mean-pooling to all objects in a frame. 
While achieving remarkable performance, we believe that we could further enhance Video QAG capability by utilizing a more fine-grained and semantic sensitive representation, such as object-level attention or contextual embedding. }
We are excited and optimistic to introduce a new perspective for targeting the challenging and practical Video QA task with cheap data. 

\section*{Acknowledgement}
This work was supported in part by the Ministry of Science and Technology, Taiwan, under Grant MOST 109-2634-F-002-032. We benefit from NVIDIA DGX-1 AI Supercomputer and are grateful to the National Center for High-performance Computing.
\ifCLASSOPTIONcaptionsoff
  \newpage
\fi



\bibliographystyle{IEEEtran}
%


\bibliography{bare_jrnl}

\begin{thebibliography}{10}
\providecommand{\url}[1]{#1}
\csname url@samestyle\endcsname
\providecommand{\newblock}{\relax}
\providecommand{\bibinfo}[2]{#2}
\providecommand{\BIBentrySTDinterwordspacing}{\spaceskip=0pt\relax}
\providecommand{\BIBentryALTinterwordstretchfactor}{4}
\providecommand{\BIBentryALTinterwordspacing}{\spaceskip=\fontdimen2\font plus
\BIBentryALTinterwordstretchfactor\fontdimen3\font minus
  \fontdimen4\font\relax}
\providecommand{\BIBforeignlanguage}[2]{{%
\expandafter\ifx\csname l@#1\endcsname\relax
\typeout{** WARNING: IEEEtran.bst: No hyphenation pattern has been}%
\typeout{** loaded for the language `#1'. Using the pattern for}%
\typeout{** the default language instead.}%
\else
\language=\csname l@#1\endcsname
\fi
#2}}
\providecommand{\BIBdecl}{\relax}
\BIBdecl

\bibitem{vdqa0}
D.~Xu, Z.~Zhao, J.~Xiao, F.~Wu, H.~Zhang, X.~He, and Y.~Zhuang, ``Video
  question answering via gradually refined attention over appearance and
  motion,'' in \emph{ACM Multimedia}, 2017.

\bibitem{vdqa1}
K.-H. Zeng, T.-H. Chen, C.-Y. Chuang, Y.-H. Liao, J.~C. Niebles, and M.~Sun,
  ``Leveraging video descriptions to learn video question answering,'' in
  \emph{AAAI}, 2017.

\bibitem{vdqa2}
Z.~Zhao, Q.~Yang, D.~Cai, X.~He, Y.~Zhuang, Z.~Zhao, Q.~Yang, D.~Cai, X.~He,
  and Y.~Zhuang, ``Video question answering via hierarchical spatio-temporal
  attention networks.'' in \emph{IJCAI}, 2017.

\bibitem{vdqa3}
Z.~Zhao, Z.~Zhang, S.~Xiao, Z.~Yu, J.~Yu, D.~Cai, F.~Wu, and Y.~Zhuang,
  ``Open-ended long-form video question answering via adaptive hierarchical
  reinforced networks.'' in \emph{IJCAI}, 2018.

\bibitem{vdqa7}
Z.~Zhao, J.~Lin, X.~Jiang, D.~Cai, X.~He, and Y.~Zhuang, ``Video question
  answering via hierarchical dual-level attention network learning,'' in
  \emph{ACM Multimedia}, 2017.

\bibitem{vdqa5}
C.~Fan, X.~Zhang, S.~Zhang, W.~Wang, C.~Zhang, and H.~Huang, ``{Heterogeneous
  Memory Enhanced Multimodal Attention Model for Video Question Answering},''
  in \emph{CVPR}, 2019.

\bibitem{vdqa6}
W.~Jin, Z.~Zhao, M.~Gu, J.~Yu, J.~Xiao, and Y.~Zhuang, ``Multi-interaction
  network with object relation for video question answering,'' in \emph{ACM
  Multimedia}, 2019.

\bibitem{vdqa8}
X.~Li, L.~Gao, X.~Wang, W.~Liu, X.~Xu, H.~T. Shen, and J.~Song, ``Learnable
  aggregating net with diversity learning for video question answering,'' in
  \emph{ACM Multimedia}, 2019.

\bibitem{vdqa9}
T.~Yang, Z.-J. Zha, H.~Xie, M.~Wang, and H.~Zhang, ``Question-aware tube-switch
  network for video question answering,'' in \emph{ACM Multimedia}, 2019.

\bibitem{vdqa10}
G.~Jiyang, G.~Runzhou, C.~Kan, and N.~Ram, ``{Motion-Appearance Co-Memory
  Networks for Video Question Answering},'' in \emph{CVPR}, 2018.

\bibitem{vdqgtmm1}
W.~{Zhang}, S.~{Tang}, Y.~{Cao}, S.~{Pu}, F.~{Wu}, and Y.~{Zhuang}, ``Frame
  augmented alternating attention network for video question answering,''
  \emph{IEEE Transactions on Multimedia}, vol.~22, no.~4, pp. 1032--1041, 2020.

\bibitem{vdqatcvst1}
Y.~{Han}, B.~{Wang}, R.~{Hong}, and F.~{Wu}, ``Movie question answering via
  textual memory and plot graph,'' \emph{IEEE Transactions on Circuits and
  Systems for Video Technology}, vol.~30, no.~3, pp. 875--887, 2020.

\bibitem{rvdqa2}
X.~Li, J.~Song, L.~Gao, X.~Liu, W.~Huang, X.~He, and C.~Gan, ``Beyond rnns:
  Positional self-attention with co-attention for video question answering,''
  in \emph{AAAI}, 2019.

\bibitem{videocap1}
Y.~Xiong, B.~Dai, and D.~Lin, ``Move forward and tell: {A} progressive
  generator of video descriptions,'' in \emph{ECCV}, 2018.

\bibitem{srcmsa}
Y.-S. Wang, H.-T. Su, C.-H. Chang, Z.-Y. Liu, and W.~Hsu, ``Video question
  generation via cross-modal self-attention networks learning,'' in
  \emph{ICASSP}, 2020.

\bibitem{anetqa}
Z.~Yu, D.~Xu, J.~Yu, T.~Yu, Z.~Zhao, Y.~Zhuang, and D.~Tao, ``Activitynet-qa: A
  dataset for understanding complex web videos via question answering,'' in
  \emph{AAAI}, 2019.

\bibitem{squad}
P.~Rajpurkar, J.~Zhang, K.~Lopyrev, and P.~Liang, ``{SQ}u{AD}: 100,000+
  questions for machine comprehension of text,'' in \emph{EMNLP}, 2016.

\bibitem{NewsQA}
A.~Trischler, T.~Wang, X.~Yuan, J.~Harris, A.~Sordoni, P.~Bachman, and
  K.~Suleman, ``Newsqa: A machine comprehension dataset,'' in \emph{Rep4NLP},
  2017.

\bibitem{tvqa}
J.~Lei, L.~Yu, M.~Bansal, and T.~Berg, ``Tvqa: Localized, compositional video
  question answering,'' in \emph{EMNLP}, 2018.

\bibitem{tgqa}
Y.~Jang, Y.~Song, Y.~Yu, Y.~Kim, and G.~Kim, ``{TGIF-QA:} toward
  spatio-temporal reasoning in visual question answering,'' in \emph{CVPR},
  2017.

\bibitem{msvd}
D.~Chen and W.~Dolan, ``Collecting highly parallel data for paraphrase
  evaluation,'' in \emph{ACL}, 2011.

\bibitem{msrvtt}
J.~Xu, T.~Mei, T.~Yao, and Y.~Rui, ``Msr-vtt: A large video description dataset
  for bridging video and language,'' in \emph{CVPR}, 2016.

\bibitem{anet}
B.~G. Fabian Caba~Heilbron, Victor~Escorcia and J.~C. Niebles, ``Activitynet: A
  large-scale video benchmark for human activity understanding,'' in
  \emph{CVPR}, 2015.

\bibitem{rvdqa1}
A.~Zadeh, M.~Chan, P.~P. Liang, E.~Tong, and L.-P. Morency, ``Social-iq: A
  question answering benchmark for artificial social intelligence,'' in
  \emph{CVPR}, 2019.

\bibitem{rvdqa3}
K.~Yi, C.~Gan, Y.~Li, P.~Kohli, J.~Wu, A.~Torralba, and J.~B. Tenenbaum,
  ``Clevrer: Collision events for video representation and reasoning,'' in
  \emph{ICLR}, 2020.

\bibitem{vqg1}
F.~Liu, T.~Xiang, T.~M. Hospedales, W.~Yang, and C.~Sun, ``Ivqa: Inverse visual
  question answering,'' in \emph{CVPR}, 2018.

\bibitem{vqg2}
Y.~Li, N.~Duan, B.~Zhou, X.~Chu, W.~Ouyang, X.~Wang, and M.~Zhou, ``Visual
  question generation as dual task of visual question answering,'' in
  \emph{CVPR}, 2018.

\bibitem{vqg3}
R.~Krishna, M.~Bernstein, and L.~Fei-Fei, ``Information maximizing visual
  question generation,'' in \emph{CVPR}, 2019.

\bibitem{cap0}
X.~Shi, J.~Cai, S.~Joty, and J.~Gu, ``Watch it twice: Video captioning with a
  refocused video encoder,'' in \emph{ACM Multimedia}, 2019.

\bibitem{cap1}
Y.~Hu, Z.~Chen, Z.-J. Zha, and F.~Wu, ``Hierarchical global-local temporal
  modeling for video captioning,'' in \emph{ACM Multimedia}, 2019.

\bibitem{cap2}
Y.~Zhu and S.~Jiang, ``Attention-based densely connected lstm for video
  captioning,'' in \emph{ACM Multimedia}, 2019.

\bibitem{cap3}
E.~Barati and X.~Chen, ``Critic-based attention network for event-based video
  captioning,'' in \emph{ACM Multimedia}, 2019.

\bibitem{cap4}
J.~Wang, W.~Wang, Y.~Huang, L.~Wang, and T.~Tan, ``Hierarchical memory
  modelling for video captioning,'' in \emph{ACM Multimedia}, 2018.

\bibitem{cap5}
S.~Liu, Z.~Ren, and J.~Yuan, ``Sibnet: Sibling convolutional encoder for video
  captioning,'' in \emph{ACM Multimedia}, 2018.

\bibitem{cap6}
H.~Wang, Y.~Xu, and Y.~Han, ``Spotting and aggregating salient regions for
  video captioning,'' in \emph{ACM Multimedia}, 2018.

\bibitem{cap7}
Z.~Yang, Y.~Han, and Z.~Wang, ``Catching the temporal regions-of-interest for
  video captioning,'' in \emph{ACMn Multimedia}, 2017.

\bibitem{cap8}
J.~Xu, T.~Yao, Y.~Zhang, and T.~Mei, ``Learning multimodal attention lstm
  networks for video captioning,'' in \emph{ACM Multimedia}, 2017.

\bibitem{cap9}
Z.~Yang, Y.~Xu, H.~Wang, B.~Wang, and Y.~Han, ``Multirate multimodal video
  captioning,'' in \emph{ACM Multimedia}, 2017.

\bibitem{cap10}
Q.~Jin, S.~Chen, J.~Chen, and A.~Hauptmann, ``Knowing yourself: Improving video
  caption via in-depth recap,'' in \emph{ACM Multimedia}, 2017.

\bibitem{cap11}
S.~Chen, J.~Chen, Q.~Jin, and A.~Hauptmann, ``Video captioning with guidance of
  multimodal latent topics,'' in \emph{ACM Multimedia}, 2017.

\bibitem{cap12}
P.~Tang, H.~Wang, H.~Wang, and K.~Xu, ``Richer semantic visual and language
  representation for video captioning,'' in \emph{ACM Multimedia}, 2017.

\bibitem{vcaptmm1}
C.~{Yan}, Y.~{Tu}, X.~{Wang}, Y.~{Zhang}, X.~{Hao}, Y.~{Zhang}, and Q.~{Dai},
  ``Stat: Spatial-temporal attention mechanism for video captioning,''
  \emph{IEEE Transactions on Multimedia}, vol.~22, no.~1, pp. 229--241, 2020.

\bibitem{vcaptcvst1}
N.~{Xu}, A.~{Liu}, Y.~{Wong}, Y.~{Zhang}, W.~{Nie}, Y.~{Su}, and
  M.~{Kankanhalli}, ``Dual-stream recurrent neural network for video
  captioning,'' \emph{IEEE Transactions on Circuits and Systems for Video
  Technology}, vol.~29, no.~8, pp. 2482--2493, 2019.

\bibitem{rvc1}
Y.~{Pan}, T.~{Mei}, T.~{Yao}, H.~{Li}, and Y.~{Rui}, ``Jointly modeling
  embedding and translation to bridge video and language,'' in \emph{CVPR},
  2016.

\bibitem{rvc2}
Y.~Pan, T.~Yao, H.~Li, and T.~Mei, ``Video captioning with transferred semantic
  attributes,'' in \emph{CVPR}, 2017.

\bibitem{rvc3}
Z.~Gan, C.~Gan, X.~He, Y.~Pu, K.~Tran, J.~Gao, L.~Carin, and L.~Deng,
  ``Semantic compositional networks for visual captioning,'' in \emph{CVPR},
  2017.

\bibitem{rvc4}
C.~Gan, Z.~Gan, X.~He, J.~Gao, and L.~Deng, ``Stylenet: Generating attractive
  visual captions with styles,'' in \emph{CVPR}, 2017.

\bibitem{rvc5}
X.~Long, C.~Gan, and G.~de~Melo, ``Video captioning with multi-faceted
  attention,'' \emph{Transactions of the Association for Computational
  Linguistics}, vol.~6, pp. 173--184, 2018.

\bibitem{resnet}
K.~He, X.~Zhang, S.~Ren, and J.~Sun, ``Deep residual learning for image
  recognition,'' in \emph{CVPR}, 2016.

\bibitem{DBLP:journals/pami/RenHG017}
S.~Ren, K.~He, R.~B. Girshick, and J.~Sun, ``Faster {R-CNN:} towards real-time
  object detection with region proposal networks,'' \emph{{IEEE} Trans. Pattern
  Anal. Mach. Intell.}, vol.~39, no.~6, pp. 1137--1149, 2017.

\bibitem{visualgenome}
R.~Krishna, Y.~Zhu, O.~Groth, J.~Johnson, K.~Hata, J.~Kravitz, S.~Chen,
  Y.~Kalantidis, L.-J. Li, D.~A. Shamma \emph{et~al.}, ``Visual genome:
  Connecting language and vision using crowdsourced dense image annotations,''
  in \emph{International journal of computer vision}, 2017.

\bibitem{transformer}
A.~Vaswani, N.~Shazeer, N.~Parmar, J.~Uszkoreit, L.~Jones, A.~N. Gomez,
  L.~Kaiser, and I.~Polosukhin, ``Attention is all you need,'' in
  \emph{{NIPS}}, 2017.

\bibitem{BLEU}
K.~Papineni, S.~Roukos, T.~Ward, and W.-J. Zhu, ``Bleu: a method for automatic
  evaluation of machine translation,'' in \emph{ACL}, 2002.

\bibitem{ROUGE}
C.-Y. Lin, ``Rouge: A package for automatic evaluation of summaries,'' in
  \emph{Text Summarization Branches Out}, 2004.

\bibitem{CIDEr}
R.~Vedantam, C.~Lawrence~Zitnick, and D.~Parikh, ``Cider: Consensus-based image
  description evaluation,'' in \emph{CVPR}, 2015.

\bibitem{meteor}
S.~Banerjee and A.~Lavie, ``{METEOR}: An automatic metric for {MT} evaluation
  with improved correlation with human judgments,'' in \emph{Intrinsic and
  Extrinsic Evaluation Measures for Machine Translation and/or Summarization},
  2005.

\bibitem{wordnet}
T.~Pedersen, S.~Patwardhan, and J.~Michelizzi, ``Wordnet::similarity: Measuring
  the relatedness of concepts,'' in \emph{NAACL}, 2004.

\bibitem{OpenNMT}
G.~Klein, Y.~Kim, Y.~Deng, J.~Senellart, and A.~M. Rush, ``Open{NMT}:
  Open-source toolkit for neural machine translation,'' in \emph{ACL}, 2017.

\bibitem{paszke2017automatic}
A.~Paszke, S.~Gross, S.~Chintala, G.~Chanan, E.~Yang, Z.~DeVito, Z.~Lin,
  A.~Desmaison, L.~Antiga, and A.~Lerer, ``Automatic differentiation in
  pytorch,'' 2017.

\bibitem{adam}
D.~P. Kingma and J.~Ba, ``Adam: {A} method for stochastic optimization,'' in
  \emph{ICLR}, 2015.

\bibitem{densecap}
L.~Zhou, Y.~Zhou, J.~J. Corso, R.~Socher, and C.~Xiong, ``End-to-end dense
  video captioning with masked transformer,'' in \emph{CVPR}, 2018.

\bibitem{unilm}
L.~Dong, N.~Yang, W.~Wang, F.~Wei, X.~Liu, Y.~Wang, J.~Gao, M.~Zhou, and H.-W.
  Hon, ``Unified language model pre-training for natural language understanding
  and generation,'' in \emph{NIPS}, 2019.

\bibitem{bert}
J.~Devlin, M.-W. Chang, K.~Lee, and K.~Toutanova, ``{BERT}: Pre-training of
  deep bidirectional transformers for language understanding,'' in
  \emph{NAACL-HLT}, 2019.

\bibitem{S2VT}
S.~Venugopalan, M.~Rohrbach, J.~Donahue, R.~Mooney, T.~Darrell, and K.~Saenko,
  ``Sequence to sequence -- video to text,'' in \emph{ICCV}, 2015.

\bibitem{obj2text}
X.~Yin and V.~Ordonez, ``Obj2text: Generating visually descriptive language
  from object layouts,'' in \emph{EMNLP}, 2017.

\bibitem{imgd}
I.~Calixto and Q.~Liu, ``Incorporating global visual features into
  attention-based neural machine translation,'' in \emph{EMNLP}, 2017.

\bibitem{VQA}
S.~Antol, A.~Agrawal, J.~Lu, M.~Mitchell, D.~Batra, C.~L. Zitnick, and
  D.~Parikh, ``{VQA}: {V}isual {Q}uestion {A}nswering,'' in \emph{ICCV}, 2015.

\bibitem{EMN}
S.~Sukhbaatar, a.~szlam, J.~Weston, and R.~Fergus, ``End-to-end memory
  networks,'' in \emph{NIPS}, 2015.

\bibitem{ESA}
L.~Yao, A.~Torabi, K.~Cho, N.~Ballas, C.~Pal, H.~Larochelle, and A.~Courville,
  ``Describing videos by exploiting temporal structure,'' in \emph{ICCV}, 2015.

\bibitem{glove}
J.~Pennington, R.~Socher, and C.~Manning, ``{G}love: Global vectors for word
  representation,'' in \emph{EMNLP}, 2014.

\end{thebibliography}
%





\end{document}